\documentclass[letterpaper, 11pt]{article}

\usepackage[utf8]{inputenc}

\usepackage{amsmath}
\usepackage{amssymb}
\usepackage{amsthm}
\usepackage[font=small]{caption} \usepackage{epsfig}
\usepackage{geometry}
\usepackage{graphicx}
\usepackage{url}
\usepackage[colorlinks=true, citecolor=blue, filecolor=black, linkcolor=red, urlcolor=blue]{hyperref}
\usepackage{multirow}
\usepackage{makecell}
\usepackage{changes}
\usepackage{algorithmic}
\usepackage[ruled]{algorithm2e}
\usepackage[caption=false]{subfig}
\usepackage{soul} \setstcolor{red}
\setulcolor{blue}
\usepackage{rotating}
\usepackage{longtable}
\usepackage[mathlines]{lineno}
\usepackage[capitalise,nameinlink,noabbrev]{cleveref} 
\usepackage{cite}

\title{Intelligent data collection for network discrimination in material flow analysis using Bayesian optimal experimental design}

\author{Jiankan Liao\footnote{\href{mailto:jkliao@umich.edu}{jkliao@umich.edu}, Ph.D. student, Mechanical Engineering, University of Michigan, Ann Arbor, MI 48109.},
Xun Huan\footnote{\href{mailto:xhuan@umich.edu}{xhuan@umich.edu}, Associate Professor, Mechanical Engineering, University of Michigan, Ann Arbor, MI 48109. \href{https://uq.engin.umich.edu}{https://uq.engin.umich.edu}}, and
Daniel Cooper\footnote{Corresponding author: \href{mailto:drcooper@umich.edu}{drcooper@umich.edu}, Associate Professor, Mechanical Engineering, University of Michigan, Ann Arbor, MI 48109.}
}

\begin{document}

\maketitle

\begin{abstract}

Material flow analyses (MFAs) are powerful tools for highlighting resource efficiency opportunities in supply chains. MFAs are often represented as directed graphs, with nodes denoting processes and edges representing mass flows. However, network structure uncertainty--uncertainty in the presence or absence of flows between nodes--is common and can compromise flow predictions. While collection of more MFA data can reduce network structure uncertainty, an intelligent data acquisition strategy is crucial to optimize the resources (person-hours and money spent on collecting and purchasing data) invested in constructing an MFA. In this study, we apply Bayesian optimal experimental design (BOED), based on the Kullback--Leibler divergence, to efficiently target high-utility MFA data---data that minimizes network structure uncertainty. We introduce a new method with reduced bias for estimating expected utility, demonstrating its superior accuracy over traditional approaches. We illustrate these advances with a case study on the U.S. steel sector MFA, where the expected utility of collecting specific single pieces of steel mass flow data aligns with the actual reduction in network structure uncertainty achieved by collecting said data from the United States Geological Survey and the World Steel Association. The results highlight that the optimal MFA data to collect depends on the total amount of data being gathered, making it sensitive to the scale of the data collection effort. Overall, our methods support intelligent data acquisition strategies, accelerating uncertainty reduction in MFAs and enhancing their utility for impact quantification and informed decision-making.

\end{abstract} 

\section{Introduction}
\label{s:intro}

Material flow analysis (MFA) is a fundamental tool in industrial ecology research for tracing the movement and transformation of resources across a supply chain. The result of an MFA is often represented as a directed graph, where nodes represent processes, materials, or locations, and edges indicate the mass flows of material. As noted by Cullen and Cooper \cite{Cullen22}, MFAs are crucial for assessing the potential environmental impacts of resource efficiency, as opportunities for improving efficiency are spread across the supply chain. 

A key challenge in constructing an MFA is the quantity and quality of available data \cite{Kopec16,Graedel19,Schwab18}, as MFA data are often:
\begin{enumerate} 
    \item \textbf{sparse}, leading to data gaps or questionable imputation from other time periods, geographies, or processes/supply chains;

    \item \textbf{noisy}, due to measurement, recording, or interpretation error; and

    \item \textbf{ill-defined}, regarding the specific section of the supply chain to which they pertain.
\end{enumerate}
Data sparsity often requires sourcing information from multiple places, further exacerbating issues with ill-defined boundaries and increasing the time and cost of constructing the MFA \cite{Schwab18}. These data challenges introduce significant epistemic uncertainty into the final MFA results due to uncertainty not only on individual mass flows between processes but also the structure of the supply chain network; i.e., the nodes and edges that define the directed graph. Uncertainty quantification (UQ) of final MFA results is gaining recognition as a critical component for ensuring study transparency and supporting informed decision-making based on MFA \cite{Cullen22,Kopec16,Schwab18}. Ideally, however, the MFA practitioner should have tools for not only \textbf{quantifying uncertainty} but also targeting \textbf{data collection} that can most efficiently reduce the uncertainty so as to allow greater confidence in the resulting metrics (e.g., recycled content) or decisions (e.g., allocation of research and development funding) informed by MFA.

\subsection{Review: Quantifying uncertainty in MFA}
\label{ss:lr_MFA_uncertainty}

Consider a mass flow, $z$, between two processes. We represent the uncertainty of this flow with a probability density function (PDF), $p(z)$. The overall uncertainty indicated by $p(z)$ can be decomposed into an MFA parametric uncertainty component (via $p(z|M_m)$ under a fixed network structure $M_m$ where $m$ indexes the $n_M$ different candidate network structures being considered) and a network structure uncertainty component (via $p(M_m)$) \cite{Liao25a}:
\begin{align}
\label{e:bayesian_averaged}
        p(z)=\sum_{m=1}^{n_M} p(z|M_m)p(M_m).
\end{align}
Below, we review the literature on quantifying both the parametric and network structure uncertainty.

\paragraph{MFA parametric uncertainty}

Addressing uncertainty in MFA results has gained increasing attention in recent years \cite{Laner14}, leading to the development of various methods to quantify it. These methods can be grouped under three banners: 1) forward error propagation \cite{Cencic08,Cencic16,Anspach24}, 2) Bayesian inference \cite{Gottschalk10,Lupton18,Dong23,Wang24,Liao25a}, and 3) determining the fit of the mass-balanced results to the data collected \cite{Kopec16,Zhu19}.

In forward error propagation, as implemented in the widely-used STAN open-source MFA package \cite{Cencic16}, collected MFA data are represented by PDFs, with at most one PDF assigned to each MFA parameter. The final mass flow uncertainty is then determined by applying mass balance constraints at each node \cite{Cencic08,Cencic15,Cencic18,Anspach24}.

Bayesian inference updates prior beliefs about MFA parameter values into posterior distributions by incorporating additional collected data, thereby reducing uncertainty. This updating process follows the rules of conditional probability \cite{Jaynes03,Berger85,Sivia06,Bertsekas08,VonToussaint11} and is particularly effective for handling sparse and noisy data. It allows the incorporation of expert domain knowledge through priors and facilitates the integration of multiple data sources. Several studies have applied Bayesian inference for UQ in MFA. Gottschalk \textit{et al.} \cite{Gottschalk10} first used this approach to quantify the uncertainty of nano-TiO2 particle releases into the environment in Switzerland. They introduced transfer coefficients (allocation fractions) to model an MFA network as a linear system using matrix algebra. Later, Lupton and Allwood \cite{Lupton18} employed a multivariate Dirichlet distribution to model the transfer coefficients emanating from a node, ensuring mass balance. They demonstrated their approach using global steel data for 2008. Dong \textit{et al.} \cite{Dong23} incorporated expert elicitation and data noise learning, deriving informed Dirichlet priors and modeling noise as a random variable. Wang \textit{et al.} \cite{Wang24} took a different approach, instead assigning priors directly on mass flows and imposing mass balance constraints via a penalty in the likelihood function, which they applied to case studies on aluminum and zinc systems.

Both the forward error propagation and Bayesian inference methods require MFA data to be modeled using PDFs. However, many data sources do not provide the necessary information to directly derive PDFs---for example, the U.N. Comtrade database's trade statistics \cite{UNcomtrade12} are not published with accompanying error bars or PDFs alongside the mass flow values. Alternatively, uncertainty in MFA can be indicated using (non)linear least squares optimization to enforce mass balance across the system, where the residuals between collected and reconciled data then serve as an indicator of how well the balanced MFA mass flow results align with the original data \cite{Kopec16, Zhu19}. However, the size of these residuals is at best indicative of uncertainty and should not be interpreted as a precise measure of mass flow uncertainty.

\paragraph{MFA network structure uncertainty}

As discussed by Liao \textit{et al.} \cite{Liao25a}, network structure uncertainty can arise from an MFA analyst's lack of subject matter expertise, outdated representations of the supply chain, or reliance on data from different regions that may not be applicable to the region of interest \cite{Klinglmair16}. Despite the ubiquitous presence of network structure uncertainty in MFA, most of the literature on MFA uncertainty has focused on quantifying mass flows (parametric) uncertainty under a fixed network structure, without considering structural uncertainty.

A few studies have explored the effects of MFA network structure. Schwab and Rechberger defined an MFA system complexity metric to describe how well the supply chain is known (0--100\%) depending on the uncertainty of the data collected and the number of nodes and edges in the network \cite{Schwab18}. Chatterjee \textit{et al.} evaluated MFA network structures using common graph-theoretic metrics as well as the concept of a ``window of vitality'' from the ecology field to assess the trade-off between supply chain resiliency and efficiency \cite{Chatterjee23}. Elsewhere, Anspach \textit{et al.}'s work touched on MFA network structure uncertainty, noting that significant discrepancies between reconciled (mass-balanced) MFA results and the original data may indicate network structure errors \cite{Anspach24}. However, to the authors' knowledge, only Liao \textit{et al.} conducted formal UQ of MFA network structures \cite{Liao25a}. They used Bayes' rule to derive the probability of different network structure candidates being the true underlying network structure as supported by the collected MFA data, demonstrating the technique with a case study on the U.S. steel sector.

\subsection{Review: Collecting MFA data}

The literature on MFA data collection focuses on acknowledging and addressing key challenges such as data sparsity, noise, and ill-defined parameters. As Graedel \cite{Graedel19} expressed, these challenges have transformed the MFA practitioner into ``\ldots part detective, part archivist, part extractor of information from experts, and part bold estimator.'' The sparsity of available data often means that MFA practitioners cannot rely solely on measurements from scientific instruments, with Brunner and Rechberger \cite{Brunner16} categorizing data collection methods into direct (e.g., through measurements) and indirect (e.g., inferred from expert estimates). While recognizing the challenges of comprehensive data collection, Khlifa \textit{et al.} \cite{Khlifa24} emphasized the importance of collecting data on as many inflows, outflows, and intermediate flows as possible, while highlighting the significant effort required to filter irrelevant data and standardize the remaining information.

Although MFA data challenges typically require discussion and analysis of uncertainty to properly contextualize the results \cite{Graedel19}, there has been limited publication on formal methods to reduce this uncertainty. One approach to reducing MFA uncertainty is through the collection of more data. Traditionally, data collection prioritization in MFA has been driven by intuition, with heuristics such as collecting all available data within resource constraints, or focusing on large mass flows that lack previous MFA estimates. While these approaches may often be adequate, there is an opportunity to enhance data acquisition based on rigorous probability and statistical frameworks. A more sophisticated, targeted approach could reduce the resources (person-hours and costs) needed to generate an MFA with sufficient certainty, ultimately enabling more informed and confident decision-making.

Under the Bayesian framework, data collection can be optimized using Bayesian optimal experimental design (BOED) \cite{Ryan16,Huan13,Chaloner95,Alexanderian21,Rainforth24,Huan24}. BOED formulates data collection as an optimization problem, where the best design option maximizes the expected utility---a measure of the experiment's value. Lindley \cite{Lindley56} introduced mutual information between model parameters and observed data (equivalently, the expected information gain in the parameters) as the expected utility, quantifying the reduction in uncertainty from prior to posterior. Liao \textit{et al.} \cite{Liao25b} recently applied BOED to MFA, identifying high-utility data sources to reduce parametric uncertainty under a fixed network structure. However, their study did not address the network structure uncertainty, which also contributes to the overall mass flow uncertainty (\cref{e:bayesian_averaged}), and is the focus of this article. 

Extending BOED to optimize data collection for reducing network structure uncertainty is a logical next step. Outside industrial ecology, BOED has been applied to model discrimination (e.g.,~\cite{Myung09,Cavagnaro10,McGree12,Drovandi14,Aggarwal15,Hainy22}), which is related to network structure discrimination in MFA. However, few of these studies explicitly estimated the expected utility based on mutual information. McGree \textit{et al.} \cite{McGree12} and Drovandi \textit{et al.} \cite{Drovandi14} used sequential Monte Carlo to characterize parameter posteriors and subsequently evaluate the model evidence (i.e., marginal likelihood). Aggarwal \textit{et al.} \cite{Aggarwal15} demonstrated BOED for model discrimination on a simple material science example where it was possible to extract the expected utility analytically. In this work, we present general numerical estimators of mutual information for model discrimination using nested Monte Carlo that can accommodate more complex, nonlinear models.

\subsection{Scope of this paper}
\label{ss:scope}

In light of the review, this study aims to define and demonstrate an intelligent, targeted MFA data collection strategy for the rapid reduction of network structure uncertainty. The key contributions of this work are: 1) applying BOED to optimize MFA data collection for reducing network structure uncertainty; 2) deriving and evaluating three numerical estimators using nested Monte Carlo sampling and solving the data collection optimization problem without the characterization of parameter posteriors; and 3) demonstrating the new framework in a case study on the U.S. steel sector. We demonstrate that the framework can effectively target the collection of single or multiple data points. 

The paper is structured as follows. \Cref{s:methodology} introduces the MFA problem formulation, the BOED framework, and the derivation of three numerical estimators for the model discrimination mutual information. A simple MFA example is provided to facilitate understanding. \Cref{s:case study} applies the framework in a case study on the 2012 U.S. steel industry. \Cref{s:discussion} discusses the trade-offs between collecting single versus multiple data points, evaluates estimator performance, and outlines the approach's limitations.
\section{Methodology}
\label{s:methodology}

The goal of intelligent data collection through BOED is to identify the data sources with the greatest potential to reduce uncertainty. To quantify this ``potential'', we first introduce the Bayesian UQ framework. Let $M_m$ represent the $m$th model (e.g., an MFA network structure) from a set of $n_M$ models, $\theta_m$ denote the parameters of that model (e.g., MFA transfer coefficients), $y$ the collected data, and $\xi$ the data collection option choice (i.e., design). Following Bayes' theorem, once $y$ is obtained from $\xi$, the uncertainty in $\theta_m$ and $M_m$ can be updated from their prior distributions to their posterior distributions:
\begin{align}
\label{e:bayes_param}
\text{(Parameter Bayes' rule)}& &   p(\theta_m|y,M_m,\xi)&=\frac{p(y|\theta_m,M_m,\xi)p(\theta_m|M_m)}{p(y|M_m,\xi)}, \\
\label{e:bayes_model}
\text{(Model Bayes' rule)}& &    p(M_m|y,\xi)&=\frac{p(y|M_m,\xi)p(M_m)}{p(y,\xi)}.
\end{align}
We refer to \cref{e:bayes_param} as the parameter Bayes' rule, where $p(\theta_m|y,M_m,\xi)$ is the parameter posterior, $p(\theta_m|M_m)$ is the parameter prior (we assume prior does not depend on the pending design $\xi$),  $p(y|\theta_m,M_m,\xi)$ is the parameter likelihood, and $p(y|M_m,\xi)$ is the model evidence (i.e., marginal likelihood). Similarly, we call \cref{e:bayes_model} the model Bayes' rule, where $p(M_m|y,\xi)$ is the model posterior, $p(M_m)$ is the model prior (similarly independent of $\xi$), and $p(y|M_m,\xi)$ (the model evidence) also appears in both equations. Notably, model evidence serves as the likelihood in \cref{e:bayes_model} and as the denominator to normalize the posterior distribution in \cref{e:bayes_param}.

With these terms established, we now define the general expected utility $U$ for collecting data from a source (i.e., design) $\xi$:
\begin{align}
\label{e:utility}
        U(\xi)&=\mathbb{E}_{y,\theta_m,m|\xi}[u(\xi,y,\theta_m,M_m)],
\end{align}
where $u(\xi,y,\theta_m,M_m)$ is a  utility function that quantifies the value of data collection under design choice $\xi$, given that the model is $M_m$, the parameters are $\theta_m$, and the observation data is $y$. Since $y$, $\theta_m$, and $M_m$ are unknown at the time of designing the experiment, we take an expectation over their joint distribution. Note that depending on the data collection options, $\xi$ may represent a single piece of data from a single source, or multiple pieces of data from the same or multiple sources.

Following an information-theoretic formulation, we define our utility function $u$ as the Kullback--Leibler (KL) divergence from the model (network structure) prior to its posterior. The KL divergence quantifies the change in distribution, where a larger value indicates a greater update in belief, and thus a more informative experiment. With this choice, the expected utility  $U$ is given by:
\begin{align}
\label{e:information_gain}
        U(\xi)=\mathbb{E}_{y|\xi}\left[D_{\text{KL}}(p_{M_m|y,\xi}\,||\,p_{M_m})\right]=\int p(y|\xi) \sum_{m=1}^{n_M} p(M_m|y,\xi) \log{\frac{p(M_m|y,\xi)}{p(M_m)}} \,\text{d}y,
\end{align}
which is equal to the mutual information between the network structure $M_m$ and the observed data $y$.

Finally, the BOED problem is to find the optimal data collection option $\xi^{\ast}$ that maximizes the expected utility:
\begin{align}
\label{e:optimal_design}
    \xi^{\ast}=\arg \max_{\xi \in \Xi} U(\xi),
\end{align}
where $\Xi$ is the set of all feasible design choices (MFA data collection options).

\begin{figure}
    \centering
    \includegraphics[width=\textwidth,keepaspectratio]{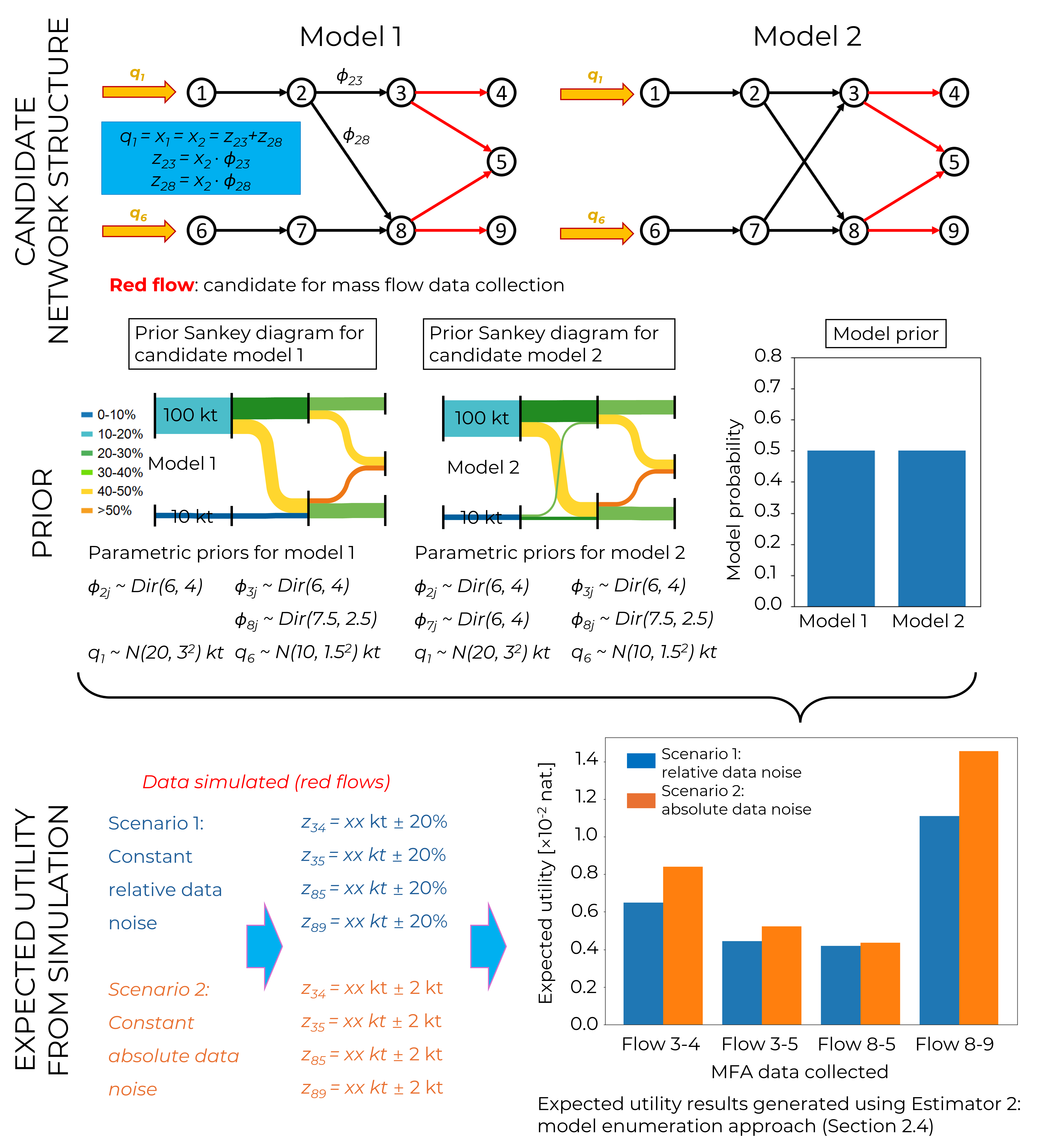}
    \caption{BOED procedure for the 9-node example MFA model. Note: $Dir(\cdot)$ represents a Dirichlet distribution. Underlying data are available at data repository: \url{https://doi.org/10.7302/k35m-xz34}
    }
    \label{f:toy_model}
\end{figure}

Solving the BOED problem requires evaluating and optimizing the expected utility $U$. However, $U$ generally lacks a closed-form solution and must be estimated numerically, requiring the computation of MFA flows under various network structures and parameter settings. 
To establish this procedure, we first represent a MFA network as a linear system in \cref{ss:MFA_framework}. We then show how to quantify MFA parametric uncertainty under a fixed network structure in \cref{ss:parametric_Bayesian} and how to quantify the MFA network structure uncertainty in \cref{ss:model_Bayesian}. Finally, in \cref{ss:BOED_framework}, we derive three different Monte Carlo sampling methods to numerically estimate the BOED expected utility in \cref{e:information_gain}. \Cref{f:toy_model} illustrates the overall BOED framework with a simple MFA example, providing a visual aid to the methodology.

\subsection{Modeling the MFA as a linear system}
\label{ss:MFA_framework}

We depict the MFA network as a directed graph where different processes, products and locations are represented as nodes (indexed $1, 2,\ldots, n_p$) and mass flows between nodes are represented as directed edges. Conservation of mass requires that at each node, the total input mass of the material flows equals the total output mass of the material flows, assuming stock accumulation or depletion is negligible. We use $x_i$ to denote the total input (equivalently, total output) flow for node $i$. The mass flow from node $i$ to $j$ is then expressed as $z_{ij}=\phi_{ij}x_i$, where $\phi_{ij} \in [0,1]$ is the allocation fraction of node $i$’s total outflow into node $j$ ($\phi_{ij}=0$ if no flow is present from node $i$ to $j$). Therefore,
\begin{align}
     \sum_{i=1}^{n_p}\phi_{ij}x_{i}+q_{j}=x_{j} \qquad \text{and} \qquad
    \sum_{j=1}^{n_p}\phi_{ij} = 1. 
\end{align}
Applying allocation fractions ($\phi_{ij}$) as model parameters instead of the direct mass flow values provides a convenient method of expressing and assembling the mass balance relationships for the entire MFA into a linear system, as explained by Gottchalk \textit{et al.} \cite{Gottschalk10}. For example, the mass balance equations for the MFA model 1 ($M_1$) illustrated in \cref{f:toy_model} (top) can be written as:
\begin{equation}
    \label{e:matrix}
    \underbrace{
        \begin{bmatrix}
            1 & 0 & 0 & 0 & 0 & 0 & 0 & 0 & 0 \\
            -\phi_{12} & 1 & 0 & 0 & 0 & 0 & 0 & 0 & 0 \\
            0 & -\phi_{23} & 1 & 0 & 0 & 0 & 0 & 0 & 0 \\
            0 & 0 & -\phi_{34} & 1 & 0 & 0 & 0 & 0 & 0 \\
            0 & 0 & -\phi_{35} & 0 & 1 & 0 & 0 & -\phi_{85} & 0 \\
            0 & 0 & 0 & 0 & 0 & 1 & 0 & 0 & 0 \\
            0 & 0 & 0 & 0 & 0 & -\phi_{67} & 1 & 0 & 0 \\
            0 & -\phi_{28} & 0 & 0 & 0 & 0 & -\phi_{78} & 1 & 0 \\
            0 & 0 & 0 & 0 & 1 & 0 & 0 & -\phi_{89} & 1
        \end{bmatrix}
    }_{\mathbb{I}-\Phi^{\top}}
    \underbrace{
        \begin{bmatrix}
            x_1 \\
            x_2 \\
            x_3 \\
            x_4 \\
            x_5 \\
            x_6 \\
            x_7 \\
            x_8 \\
            x_9
        \end{bmatrix}
    }_{x}
    =
    \underbrace{
        \begin{bmatrix}
            q_1 \\
            0 \\
            0 \\
            0 \\
            0 \\
            q_6 \\
            0 \\
            0 \\
            0
        \end{bmatrix}
    }_{q},
\end{equation}
where $\mathbb{I}$ is the $n_p \times n_p$ identity matrix and $\Phi \in \mathbb{R}^{n_p \times n_p}$ is the adjacency matrix with entries representing the allocation fractions $\phi_{ij}$. $x \in \mathbb{R}^{n_p}$ and $q \in \mathbb{R}^{n_p}$ are column vectors representing all nodal mass flows and external inflows to the network (see \cref{f:toy_model}, top), respectively. The material inflows ($q_i$) are from outside the network; e.g., aluminum imports in a country-level aluminum MFA.

With $\Phi$ and $q$ established, all nodal mass flows $x$ can be solved via:
\begin{equation}
\label{e:node}
    x=(\mathbb{I}-\Phi^{\top})^{-1}q.
\end{equation}
The transformation of the allocation fraction matrix $(\mathbb{I}-\Phi)^{-1}$ is also known as the Ghosh inverse~\cite{Ghosh58}, serving as a supply-driven alternative to the more common demand-driven input/output (I/O) analysis~\cite{Leontief36}. Other common MFA quantities of interest (QoIs) can be derived from the values for ${x,\Phi}$ and $q$; e.g., mass flows for each connection, $z_{ij}=\phi_{ij}x_i$. We represent these QoIs through a vector-valued function, $G(\Phi,q,\xi;M_m)$, which typically corresponds to the same quantities as those in the collected MFA data, $y$.

For each network structure $M_m$, we designate $\theta_m=\{\phi_{ij},q_i | \phi_{ij}\neq \text{const.},q_i\neq \text{const.} \,\text{under}\, M_m\}$ to describe the set of all uncertain model parameters (i.e., of existing connections and external inflows) under said network structure. The predicted QoIs can then be written as $G(\theta_m,\xi;M_m)$. As an example, in \cref{f:toy_model} (top), mass flow $z_{73}$ is not present in the network structure $M_1$ but is in $M_2$, then $\phi_{73}$ is a trivial parameter always equal to zero and therefore not included in $\theta_{m=1}$, but would be a non-trivial parameter and therefore part of $\theta_{m=2}$. Hence, the composition of $\theta_m$ varies depending on the network structure $M_m$, and the subscript $m$ concisely captures this distinction.

\subsection{Parametric uncertainty under a fixed network structure}
\label{ss:parametric_Bayesian}

We now outline our specific prior and likelihood setup in the parameter Bayes' rule (\cref{e:bayes_param}) for capturing parametric uncertainty. To ensure mass balance constraints are automatically satisfied, we assign Dirichlet priors to the allocation fractions $\phi$. For mass flow inputs $q$, we use truncated normal priors with a non-negative lower bound. Prior hyper-parameters can be determined through expert elicitation \cite{Dong23}, leveraging domain expertise to reduce the subsequent volume of data that must be collected to reach a desired reduction in uncertainty. Alternatively, they can be specified using historical data or by adopting non-informative priors~\cite{Lupton18} when domain knowledge is limited.

The likelihood quantifies the probability of observing data $y$, given the model parameters $\theta_m$, network structure $M_m$, and the data collection option (design) $\xi$. It provides a probabilistic measure of the mismatch between the observation $y$ and the model predicted QoI, $G(\theta_m,\xi;M_m)$. One approach is to form the likelihood through an additive noise data model~\cite{Dong23}. However, in this study, we adopt a relative noise model to better capture proportional noise in the data:
\begin{equation}
\label{e:relative_error}
    y_k=G_k(\theta_m,\xi;M_m)(1+\epsilon_k),
\end{equation}
where the subscript $k$ indicates the $k$th component of the vector, and $\epsilon_k\sim \mathcal{N}(0,\sigma_k^2)$ is an independent relative noise term. Subsequently, we obtain
\begin{align}
\label{e:likelihood}
       p(y|\theta_m,M_m) & =\prod_{k=1}^{n_y}p_{\epsilon_k}\left(\frac{y_k}{G_k(\theta_m,\xi;M_m)}-1\right) \left|\frac{\text{d}}{\text{d}y_k}\left(\frac{y_k}{G_k(\theta_m,\xi;M_m)}-1\right)\right| \nonumber\\
       & =\prod_{k=1}^{n_y}\frac{1}{\sqrt{2\pi}\sigma_k}\exp\left[-\frac{1}{2\sigma_k^2}\left(\frac{y_k}{G_k(\theta_m,\xi;M_m)}-1\right)^2\right] \frac{1}{G_k(\theta_m,\xi;M_m)},
\end{align}
as a result of the independent $\epsilon_k$'s.
We can assign a fixed value to the data standard deviation (e.g., $\sigma_k=0.1$, corresponding to a level of $\pm 10\%$ relative noise). Alternatively, the noise term can be modeled dynamically using different approaches. One option is to treat it as an inferable parameter within the model parameter set $\theta_m$ \cite{Dong23}. Another approach is to manually adjust the noise value and repeatedly solve the inference problem until the Bayesian posterior predictive aligns well with the collected data \cite{Wang24}. Both such approaches, however, would require substantially higher computational cost for solving the overall Bayesian inference problem.

\subsection{Network structure uncertainty}
\label{ss:model_Bayesian}

We now outline our specific prior and likelihood setup in the model Bayes' rule (\cref{e:bayes_model}) for capturing network structure uncertainty. First, we need to generate a pool of $n_M$ candidate node-and-flow network structures. We restrict this uncertainty to the existence of connections between nodes. These candidate network structures may already be known to the MFA analyst. Otherwise, as detailed by Liao \textit{et al.} \cite{Liao25a}, a combination of \emph{exploitation} and \emph{exploration} can be adopted to populate the candidate pool:  identifying a series of uncertain connections by comparing network structures used in existing MFAs on the same or similar subjects or consulting domain experts (exploitation), and generating semi-randomized ``wild-guesses'' about the existence or absence of certain connections to enhance the diversity of the candidate pool (exploration). Once a total number of $n_L$ critical uncertain connections has been identified, a complete permutation of these connections yields $2^{n_L}$ possible candidate network structures. After establishing the candidate network structures, a model prior distribution is assigned for the network structures, $p(M_m)$. Liao \textit{et al.} \cite{Liao25a} adopted an informative prior based on expert elicitation of the probability that individual uncertain connections exist. Alternatively, an uninformed uniform prior can be used, setting $p(M_m)=\frac{1}{n_M}$ for all $m$.

The model likelihood, $p(y|M_m,\xi)$, is in fact the model evidence (denominator) in the parameter Bayes' rule (\cref{e:bayes_param}). It can be explicitly expressed as an integral over the parameter likelihood:
\begin{align}
\label{e:model_likelihood}
    p(y|M_m,\xi)=\int p(y|\theta_m,M_m,\xi)p(\theta_m|M_m) \, \text{d}\theta_m.
\end{align}

\subsection{Mutual information estimators for Bayesian model selection in MFA}
\label{ss:BOED_framework}

We now derive three numerical estimators for the mutual information expected utility in \cref{e:information_gain}, following a nested Monte Carlo structure that avoids the need for explicit evaluation or sampling from the parameter posteriors. The key difference between these estimators is their sampling processes in the outer loop. To reflect these differences, we refer to the three estimators as: 1) \emph{data-model joint MC}, 2) \emph{model enumeration}, and 3) \emph{data marginal MC}.

\paragraph{Estimator 1: data-model joint MC}

We begin by applying Bayes' rule to \cref{e:information_gain} and then use a standard Monte Carlo approximation: 
\begin{align}
    U(\xi) 
    &=\int p(y|\xi) \sum_{m=1}^{n_M} p(M_m|y,\xi) \log{\frac{p(M_m|y,\xi)}{p(M_m)}} \,\text{d}y \nonumber\\
    &=\int \sum_{m=1}^{n_M} p(M_m)p(y|M_m,\xi) \log{\frac{p(y|M_m,\xi)}{p(y|\xi)}} \,\text{d}y \nonumber\\
    & \approx \frac{1}{N_{\text{out}}}\sum_{\ell_1=1}^{N_{\text{out}}} \left[\log{p(y^{(\ell_1)}|M_{m^{(\ell_1)}},\xi)-\log{p(y^{(\ell_1)}|\xi)}}\right],
    \label{e:est1_outer}
\end{align}
where $N_{\text{out}}$ is the number of Monte Carlo samples, and samples $M_{m^{(\ell_1)}}\sim p(M_m)$ are drawn from the model prior and $y^{(\ell_1)}\sim p(y|M_{m^{(\ell_1)}},\xi)$ from the model evidence (by first drawing a temporary parameter sample $\theta'\sim p(\theta)$ followed by $y^{(\ell_1)}\sim p(y|\theta',M_{m^{(\ell_1)}},\xi)$). 
However, the two log-density terms in \cref{e:est1_outer} are not available analytically. The first term, as defined in \cref{e:model_likelihood}, can be approximated using an inner-loop Monte Carlo:
\begin{align}
\log p(y^{(\ell_1)}|M_{m^{(\ell_1)}},\xi) &= \log\left( \int p(y^{(\ell_1)}|\theta_{m^{(\ell_1)}},M_{m^{(\ell_1)}},\xi) p(\theta_{m^{(\ell_1)}}|M_{m^{(\ell_1)}}) \, \text{d}\theta_{m^{(\ell_1)}} \right) \nonumber\\
&\approx \log \left(\frac{1}{N_{\text{in},1}}\sum_{\ell_2=1}^{N_{\text{in},1}} p(y^{(\ell_1)}|\theta_{m^{(\ell_1)}}^{(\ell_2)},M_{m^{(\ell_1)}},\xi)\right),
\label{e:est1_term1}
\end{align}
where $N_{\text{in},1}$ is the number of Monte Carlo samples, $\theta_{m^{(\ell_1)}}$ represents the parameter random vector corresponding to the model whose index has been sampled to be $m^{(\ell_1)}$ from the outer loop, and $\theta_{m^{(\ell_1)}}^{(\ell_2)}$ denotes the $\ell_2$-th sample of that parameter random vector drawn from its prior distribution $p(\theta_{m^{(\ell_1)}}|M_{m^{(\ell_1)}})$.
The second term can be similarly approximated using an inner-loop Monte Carlo:
\begin{align}
\log p(y^{(\ell_1)}|\xi) &= \log\left( \sum_{m=1}^{n_M} \int p(y^{(\ell_1)}|\theta_{m},M_{m},\xi) p(\theta_{m}|M_{m}) p(M_m) \, \text{d}\theta_{m} \right) \nonumber\\
&\approx \log \left(\sum_{m=1}^{n_M} \left[\frac{1}{N_{\text{in},2}}\sum_{\ell_3=1}^{N_{\text{in},2}} p(y^{(\ell_1)}|\theta_{m}^{(\ell_3)},M_{m},\xi) \right] p(M_m) \right),
\label{e:est1_term2}
\end{align}
where $N_{\text{in},2}$ is the number of Monte Carlo samples, and samples $\theta_{m}^{(\ell_3)}\sim p(\theta_{m}|M_{m})$ are drawn from its prior distribution. 
Note that the $\theta_m$ and $M_m$ terms are ``internal'' variables of marginalization and do not depend on the outer-loop $\ell_1$ sample.
Only the $\theta_m$ marginalization is approximated by Monte Carlo, while that for $M_m$ is summed exactly over all $n_M$ models (this can also be replaced by a Monte Carlo).
Additionally, the two inner loops may, but do not have to, have the same number sample sizes, $N_{\text{in},1}$ and $N_{\text{in},2}$. 
Finally, combining \cref{e:est1_term1,e:est1_term2} into \cref{e:est1_outer}, we obtain the overall estimator 1, $\widehat{U}_1$:
\begin{align}
    U(\xi) \approx \widehat{U}_1(\xi) &:=
    \frac{1}{N_{\text{out}}}\sum_{\ell_1=1}^{N_{\text{out}}} \left[
    \log \left(\frac{1}{N_{\text{in},1}}\sum_{\ell_2=1}^{N_{\text{in},1}} p(y^{(\ell_1)}|\theta_{m^{(\ell_1)}}^{(\ell_2)},M_{m^{(\ell_1)}},\xi)\right)
    \right. \nonumber\\
    &\hspace{5.4em} \left.
    -\log \left(\sum_{m=1}^{n_M} \left[\frac{1}{N_{\text{in},2}}\sum_{\ell_3=1}^{N_{\text{in},2}} p(y^{(\ell_1)}|\theta_{m}^{(\ell_3)},M_{m},\xi) \right] p(M_m) \right)\right].
    \label{e:numerical_utility_1}
\end{align}

\paragraph{Estimator 2: model enumeration}

The second estimator differs from $\widehat{U}_1$ by exhaustively looping over all $n_M$ candidate network structures in the outer loop instead of randomly sampling. This eliminates the additional variation from Monte Carlo but at the cost of needing to complete looping over all network structures. \Cref{e:est1_outer} then becomes:
\begin{align}
    U(\xi) 
    &=\int p(y|\xi) \sum_{m=1}^{n_M} p(M_m|y,\xi) \log{\frac{p(M_m|y,\xi)}{p(M_m)}} \,\text{d}y \nonumber\\
    &=\sum_{m=1}^{n_M}\int p(M_m)  p(y|M_m,\xi) \log{\frac{p(y|M_m,\xi)}{p(y|\xi)}} \,\text{d}y \nonumber\\
    & \approx \sum_{m=1}^{n_M}\frac{1}{N_{\text{out}}}\sum_{\ell_1=1}^{N_{\text{out}}} p(M_m)\left[\log{p(y^{(\ell_1)}|M_{m},\xi)-\log{p(y^{(\ell_1)}|\xi)}}\right],
    \label{e:est2_outer}
\end{align}
where $y^{(\ell_1)}\sim p(y|M_m,\xi)$; note that $M_m$ is no longer a sample, but iterated over the index $m$ in the summation. Following a similar derivation as estimator 1, we arrive at the overall estimator 2, $\widehat{U}_2$: 
\begin{align}
    U(\xi) \approx \widehat{U}_2(\xi) &:=
    \sum_{m_1=1}^{n_M} p(M_{m_1}) \frac{1}{N_{\text{out}}}\sum_{\ell_1=1}^{N_{\text{out}}} \left[
    \log \left(\frac{1}{N_{\text{in},1}}\sum_{\ell_2=1}^{N_{\text{in},1}} p(y^{(\ell_1)}|\theta_{m_1}^{(\ell_2)},M_{m_1},\xi)\right)
    \right. \nonumber\\
    &\hspace{9em} \left.
    -\log \left(\sum_{m_2=1}^{n_M} \left[\frac{1}{N_{\text{in},2}}\sum_{\ell_3=1}^{N_{\text{in},2}} p(y^{(\ell_1)}|\theta_{m_2}^{(\ell_3)},M_{m_2},\xi) \right] p(M_{m_2}) \right)\right].
    \label{e:numerical_utility_2}
\end{align}

\paragraph{Estimator 3: data marginal MC}

The third estimator is formed by interpreting \cref{e:information_gain} as varying KL divergence realizations (of the model prior to posterior) under different possible data observations:
\begin{align}
    U(\xi) 
    &=\int p(y|\xi) \sum_{m=1}^{n_M} p(M_m|y,\xi) \log{\frac{p(M_m|y,\xi)}{p(M_m)}} \,\text{d}y \nonumber\\
    &=\int p(y|\xi) \sum_{m=1}^{n_M} \frac{p(y|M_m,\xi) p(M_m)}{p(y|\xi)} \log{\frac{p(y|M_m,\xi)}{p(y|\xi)}} \,\text{d}y \nonumber\\
    & \approx \frac{1}{N_{\text{out}}}\sum_{\ell_1=1}^{N_{\text{out}}} \sum_{m=1}^{n_M} \frac{p(y^{(\ell_1)}|M_m,\xi) p(M_m)}{p(y^{(\ell_1)}|\xi)}
    \left[\log{p(y^{(\ell_1)}|M_m,\xi)-\log{p(y^{(\ell_1)}|\xi)}}\right],
    \label{e:est3_outer}
\end{align}
where $y^{(\ell_1)} \sim p(y|\xi)$ are samples drawn from the marginal distribution for $y$ (by first drawing a temporary model sample $M_{m'} \sim p(M_m)$, a temporary parameter sample $\theta'\sim p(\theta)$, and then $y^{(\ell_1)} \sim p(y|\theta',M_{m'},\xi)$).
The two terms, $p(y^{(\ell_1)}|M_m,\xi)$ and $p(y^{(\ell_1)}|\xi)$, appear twice in \cref{e:est3_outer}---once outside the logarithm and once inside---and need to be approximated separately using, e.g., the inner-loop Monte Carlo formulas from \cref{e:est1_term1,e:est1_term2}. Substituting them in, we arrive at the overall estimator 3, $\widehat{U}_3$:
\begin{align}
    U(\xi) \approx \widehat{U}_3(\xi) &:=
    \frac{1}{N_{\text{out}}}\sum_{\ell_1=1}^{N_{\text{out}}} \sum_{m_1=1}^{n_M}
    \left\{\frac{\left[\frac{1}{N_{\text{in},1}}\sum_{\ell_2=1}^{N_{\text{in},1}} p(y^{(\ell_1)}|\theta_{m_1}^{(\ell_2)},M_{m_1},\xi)\right]p(M_{m_1})}{\sum_{m_2=1}^{n_M} \left[\frac{1}{N_{\text{in},2}}\sum_{\ell_3=1}^{N_{\text{in},2}} p(y^{(\ell_1)}|\theta_{m_2}^{(\ell_3)},M_{m_2},\xi) \right] p(M_{m_2})} \right. \nonumber\\
    &\hspace{7.4em} \left[ \log \left(\frac{1}{N_{\text{in},1}}\sum_{\ell_2=1}^{N_{\text{in},1}} p(y^{(\ell_1)}|\theta_{m_1}^{(\ell_2)},M_{m_1},\xi)\right)
    \right. \nonumber\\
    &\hspace{7em} \left.\left.
    -\log \left(\sum_{m_2=1}^{n_M} \left[\frac{1}{N_{\text{in},2}}\sum_{\ell_3=1}^{N_{\text{in},2}} p(y^{(\ell_1)}|\theta_{m_2}^{(\ell_3)},M_{m_2},\xi) \right] p(M_{m_2}) \right)\right]\right\}.
    \label{e:numerical_utility_3}
\end{align}

To reduce the computational cost of the nested Monte Carlo estimators, we implement sample reuse techniques, which also helps prevent near-zero model evidence estimates at small sample sizes \cite{Huan13}. This approach involves generating a batch of prior samples and pre-computing their corresponding $G(\theta_m,\xi;M_m)$ for all $n_M$ candidate network structures. These pre-computed values are then reused for both the outer and inner loops
(e.g., for $\widehat{U}_1$, setting $\theta'=\theta^{(\ell_2)}_{m^{(\ell_1)}}=\theta^{(\ell_3)}_m$ and $N_{\text{out}}=N_{\text{in},1}=N_{\text{in},2}$). 
While sample reuse may introduce some bias in the estimators, the effect is very small \cite{Huan10}. Notably, sample reuse significantly reduces the computational complexity in terms of unique likelihood evaluations (i.e., unique MFA solves) at each $\xi$. Specifically, the cost is reduced from $\mathcal{O}\left(N_{\text{out}}(N_{\text{in},1}+n_MN_{\text{in},2})\right)$ to $\mathcal{O}\left(n_M N_{\text{out}}\right)$ for $\widehat{U}_1$, from $\mathcal{O}\left(n_M N_{\text{out}}(N_{\text{in},1}+n_MN_{\text{in},2})\right)$ to $\mathcal{O}\left(n_MN_{\text{out}}\right)$ for $\widehat{U}_2$, and from $\mathcal{O}\left(n_M N_{\text{out}}(N_{\text{in},1}+n_MN_{\text{in},2})\right)$ to $\mathcal{O}\left(n_MN_{\text{out}}\right)$ for $\widehat{U}_3$.

The simple MFA example in \cref{f:toy_model} illustrates the BOED results using $\widehat{U}_2$. Among the four available downstream mass flow data collection options, BOED identifies measuring the mass flow on $z_{89}$ as the optimal choice for reducing network structure uncertainty. This result aligns with intuition, as the existence/absence of the mass flow $z_{73}$ would cause a larger percentage fluctuation on node 8, which has a smaller nodal mass flow than node 3. In addition, mass flow $z_{89}$ receives a larger allocation out of the flows originating from node 8. Consequently, collecting mass flow data on $z_{89}$ is more likely to reveal the underlying network structure than the other data collection options.
\section{Case study on the U.S. steel flow}
\label{s:case study}

We demonstrate the BOED framework for reducing MFA network structure uncertainty through a case study on the U.S. steel flow in 2012. We select this year to maintain consistency with previous work on uncertainty quantification and reduction in MFA using Bayesian inference \cite{Dong23, Liao25a, Liao25b}, though the framework is applicable to any year. All data and code for this case study are available online (see the SI). 

\subsection{Candidate network structures and prior distributions}
\label{ss:cs_model}
\begin{figure}
    \centering
    \includegraphics[height=0.9\textheight,keepaspectratio]{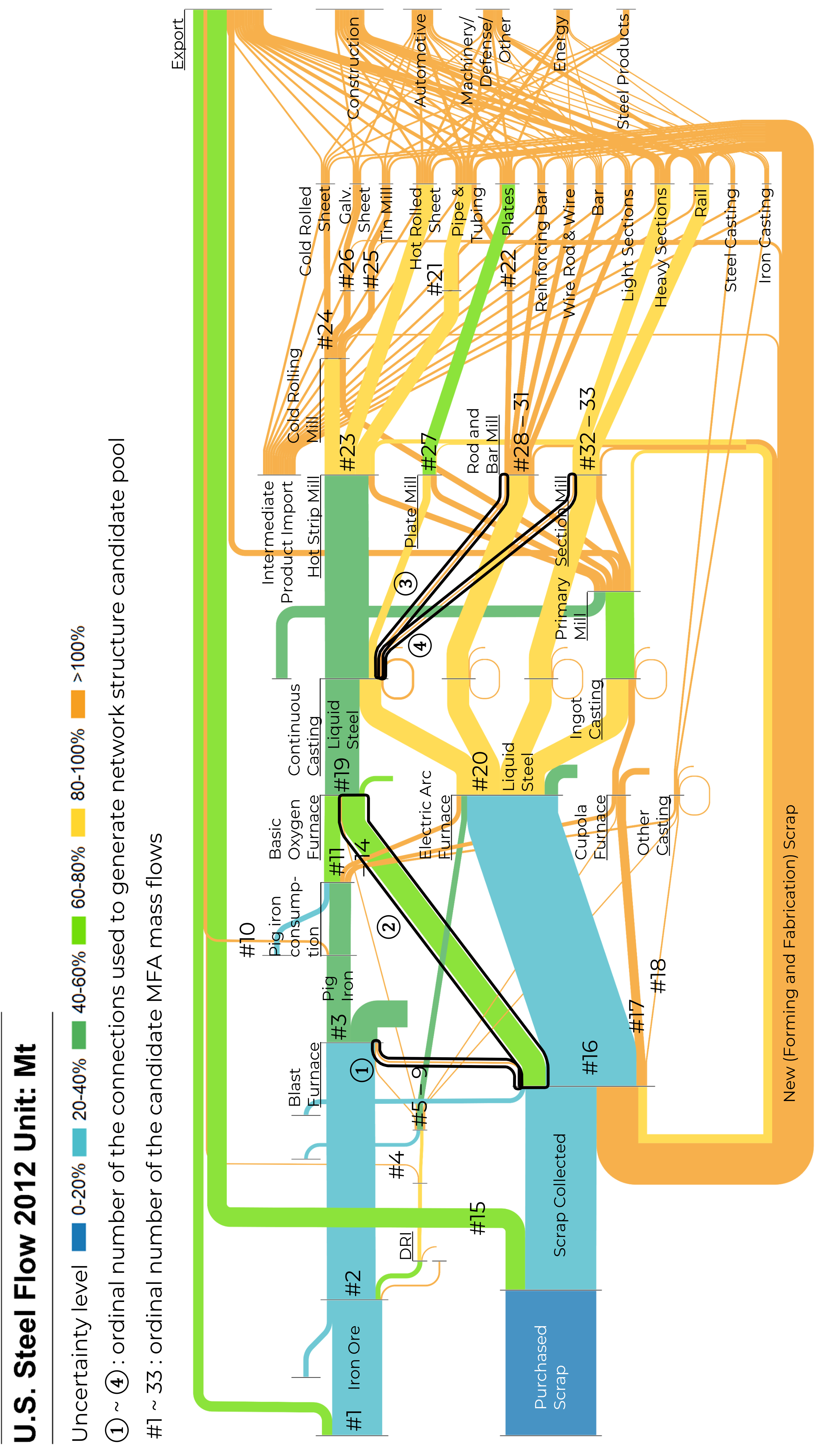}
    \caption{Bayesian prior-predictive mass flows for the U.S. steel flow in 2012.
    The uncertainty percentages refer to the flow standard deviation as a percentage of the mean of the mass flow. All mass flows refer to steel except for the iron ore flows that include the non-iron mass (e.g., oxygen and gangue). 
    Candidate MFA mass flows for data collection are marked in the ``\#'' flows with numbering (see Table S1 
    for details) Underlying data are available at data repository: \url{https://doi.org/10.7302/k35m-xz34}
    }
    \label{f:sankey_diagram}
\end{figure}

We revisit the 16 candidate network structures for the U.S. steel sector defined by Liao \textit{et al.} \cite{Liao25a}. They extracted a baseline network structure from Zhu \textit{et al.} \cite{Zhu19}, which includes 270 metal flows (edges) connecting 55 nodes. Based on the exploitation approach (discussions with industry experts) and the exploration approach (semi-randomized guesses), Liao \textit{et al.} identified four edges whose presence or absence in the network was uncertain: (1) post-consumer steel scrap to the blast furnace (BF, connection index 1); (2) scrap to the basic oxygen furnace (BOF, connection index 2); (3) the BOF continuously cast slab (CC) to the rod and bar mill (RBM, connection index 3); and (4) the BOF CC to the section mill (connection index 4). These edges are highlighted in the Sankey diagram in \cref{f:sankey_diagram}. The 16 candidate network structures were generated from the complete permutation of the 4 questionable flows ($2^4=16$). Per Liao \textit{et al.}, the network structures are described using a 4-digit binary code to indicate whether the indexed connection is present (1) or absent (0). For example, none of the 4 targeted connections exist in the network structure, 0000; whereas, only the flow between scrap and the blast furnace (index 1) exists in the network structure, 1000. A uniform prior is used to model the network structure uncertainty before any data is collected, with the probability for each network structure equal to $\frac{1}{16}$. 

For each candidate network structure, we use the same parameter priors as Liao \textit{et al.} They formed informative parameter priors for upstream allocation fractions ($\phi_{ij}$'s) and external inputs ($q_i$'s) using expert elicitation, and used non-informative priors for the downstream allocation fractions. These prior distributions are provided in S1 
and readers are directed to Dong \textit{et al.} \cite{Dong23} for the methodology to conduct expert elicitation. \Cref{f:sankey_diagram} shows the Sankey diagram representation of the U.S. steel sector MFA using the parameter priors and a network structure of 1111.

\subsection{Targeting MFA data collection}
\label{ss:cs_data_collection}

A total of 33 candidates have been identified for data collection. Each candidate is an absolute mass flow value published by the United States Geological Survey (USGS) \cite{USGS12a, USGS12b, USGS12c} or the World Steel Association (WSA) \cite{WSA12}. The 33 candidates for mass flow data collection are labeled with ``\#'' numbers in \cref{f:sankey_diagram}, with the immediate upstream and downstream nodes defined in Table S1 of the SI. 
Per the work of Lupton and Allwood \cite{Lupton18} and Dong \textit{et al.} \cite{Dong23} on steel flow analyses, all candidates for data collection are assumed to exhibit a relative noise (standard deviation of a normal distribution) of 10\% of the nominal value.
Using the BOED framework (see \cref{ss:BOED_framework}), we target the collection of data that reduces the network structure uncertainty when collecting a single piece of data and when collecting multiple pieces of data.

\subsection{Case study results}
\label{ss:cs_result}

\subsubsection{Expected utility results for single-piece data record collection}
\label{sss:single_utility}

We use all three mutual information estimators proposed in \cref{ss:BOED_framework} to solve the BOED problem for the case study on the U.S. steel flow with 16 candidate network structures (each containing approximately 180 parameters). 
Using the sample reuse technique outlined in \cref{ss:BOED_framework}, we allocate 160,000 unique MFA solves to each mass flow data collection option, $\xi$, across all three methods. This configuration sets parameter sample sizes to 10,000 for each estimator: 160,000 unique evaluations divided by the 16 candidate network structures. It took 30 seconds to compute $\widehat{U}_1$ ({data-model joint MC} estimator) and $\widehat{U}_3$ ({data marginal MC} estimator) and 10 minutes for $\widehat{U}_2$ ({model enumeration} estimator) using an Intel(R) Core i7-9700K CPU, 3.60GHz. 

\begin{figure}
    \centering
    \includegraphics[width=0.7\textwidth,keepaspectratio]{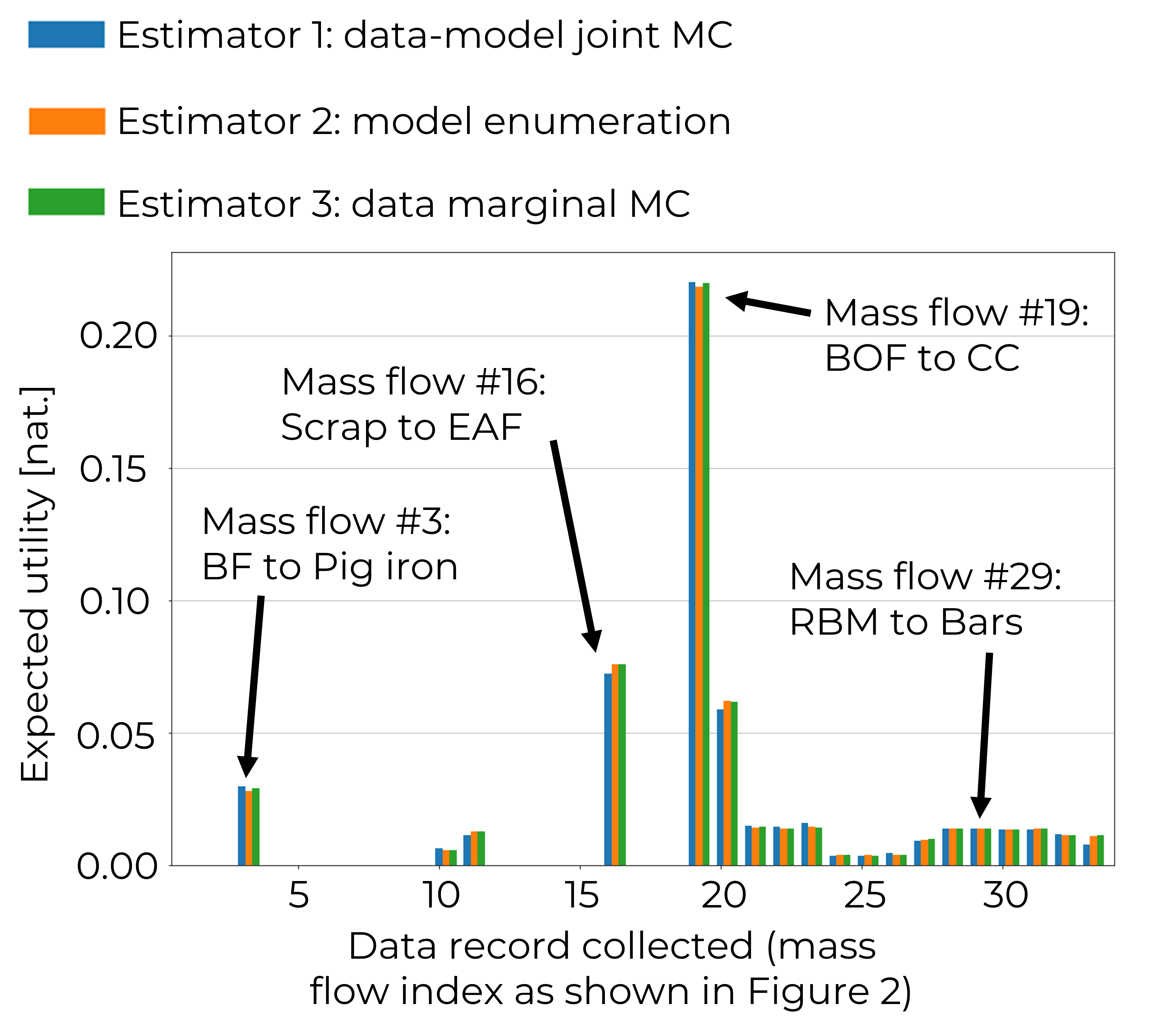}
    \caption{Expected utility of collecting a single data record on U.S. steel sector mass flows to reduce MFA network structure uncertainty. The expected utility is evaluated using the three numerical estimators derived in \cref{ss:BOED_framework}. Underlying data are available at data repository: \url{https://doi.org/10.7302/k35m-xz34}
    }
    \label{f:expected utility}
\end{figure}

\Cref{f:expected utility} presents the expected utility for all 33 mass flow data collection designs, evaluated using all three estimators. The figure shows that collecting data on mass flow \#19 (the output of the BOF to continuous casting) has the highest expected utility, regardless of the estimator used. The second most valuable piece of data is expected to be mass flow \#16, the flow of scrap into the electric arc furnace (EAF). Upon examining \cref{f:sankey_diagram}, it is evident that collecting data on mass flows \#16 and \#19 are likely valuable for reducing network structure uncertainty because they share nodes with multiple uncertain flows: mass flow \#16 shares nodes with connection indices 1 and 2, while mass flow \#19 shares nodes with indices 2, 3, and 4. In contrast, collecting data on many other mass flows is expected to have negligible utility in reducing uncertainty. For example, collecting data on mass flows \#5-9, which represent relatively minor upstream flows directing Direct Reduced Iron (DRI) to different furnaces, is expected to contribute little to reducing uncertainty. While collecting data on other mass flows, such as \#3 and \#29, is still expected to be useful for reducing network structure uncertainty, it is not as expected to be as effective as collecting data on mass flows \#16 and \#19. Referring to \cref{f:sankey_diagram}, mass flows \#3 and \#29 each share a node with only one uncertain connection, limiting their impact on uncertainty reduction compared to flows \#16 and \#19.

\begin{figure}
    \centering
    \includegraphics[width=\textwidth,keepaspectratio]{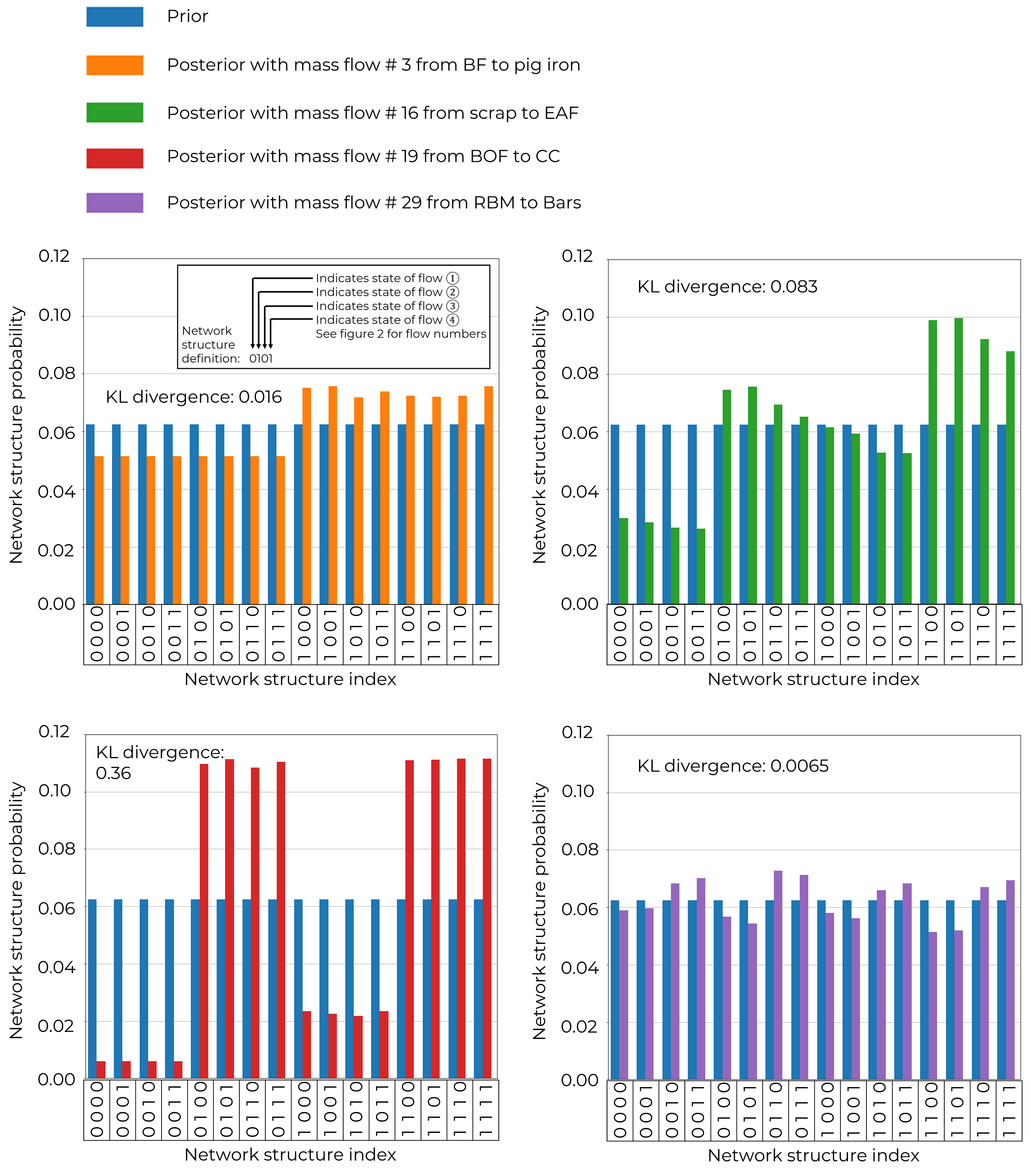}
    \caption{Prior and posterior probability for all 16 candidate network structures with data collection on selected mass flow candidates. Underlying data are available at data repository: \url{https://doi.org/10.7302/k35m-xz34}
    }
    \label{f:BOED validation}
\end{figure}

To test the alignment between the expected utility results and the actual reduction in network structure uncertainty achieved by data collection, we collect data on mass flows \#3, \#16, \#19, and \#29 from the USGS \cite{USGS12a, USGS12b, USGS12c} and WSA \cite{WSA12} (see Table S1). 
We then update the network structure uncertainty using \cref{e:bayes_model}. \Cref{f:BOED validation} presents the network structure prior and posterior distributions, along with the KL divergence from prior to posterior after collecting data on the selected mass flows. A higher KL divergence indicates a greater change from the prior and posterior, reflecting a larger reduction in uncertainty. The results show alignment between the order of the expected utility values for these flows (from highest to lowest: \#19, \#16, \#3, \#29), as shown in \Cref{f:expected utility}, and the actual utility achieved by collecting the data, as demonstrated in \Cref{f:BOED validation}.

\subsubsection{Expected utility result for multi-piece data record collection}
\label{sss:batch_utility}

The updated MFA uncertainty from collecting a single-piece of MFA data could be the starting point for a new round of BOED and collection of a second data point. An alternative to this one-at-a-time approach to data collection is to use BOED to target the collection of batches of data. To demonstrate this, the BOED framework was used to target optimal collection of two-pieces of MFA data, where we explore all the possible combinations to collect two pieces of data ($33\times 17=561$ total combinations).
The same settings for the estimators are chosen for this analysis where we allocate a total of 160,000 unique MFA solves to each mass flow data collection option with sample reuse.

\begin{figure}
    \centering
    \includegraphics[width=0.8\linewidth]{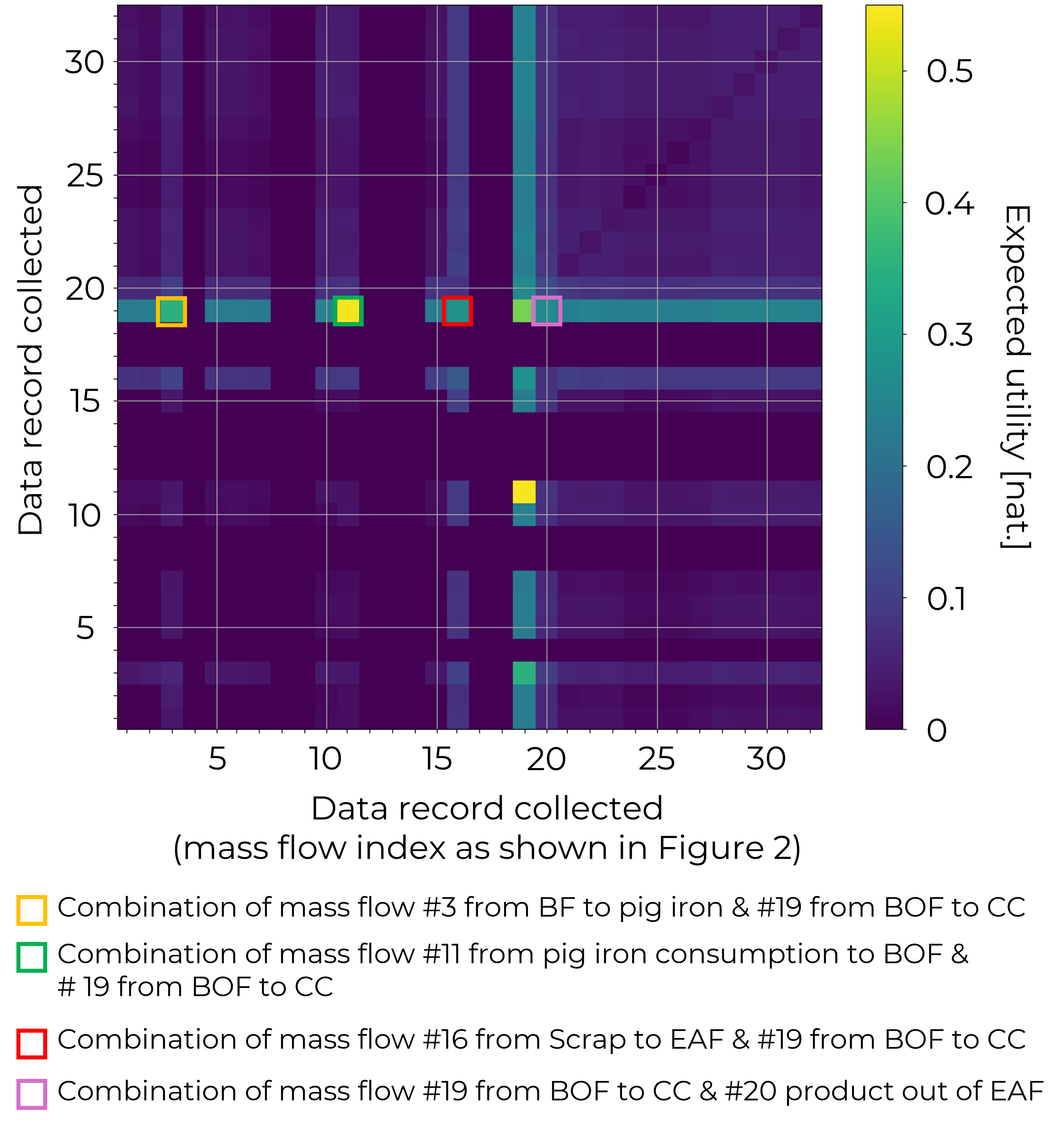}
    \caption{Expected utility of multi-piece data collection - collecting two mass flow data records at the same time. The expected utility is evaluated using Estimator 2): \emph{model enumeration} approach, derived in \cref{ss:BOED_framework}. Underlying data are available at data repository: \url{https://doi.org/10.7302/k35m-xz34}
    }
    \label{f:batch_utility}
\end{figure}

\Cref{f:batch_utility} presents the expected utility for all data collection combinations and is symmetric along the diagonal line, $y=x$. The highest expected utility is achieved by collecting data on both mass flow \#11 (from Pig iron consumption to BOF) and mass flow \#19 (output of BOF to continuous casting). Other high-utility combinations include mass flows \#3 and \#19, \#16 and \#19, and \#19 and \#20. It is also important to note that collecting independent data on mass flow \#19 twice is expected to yield a high utility.

\begin{figure}
    \centering
    \includegraphics[width=\linewidth]{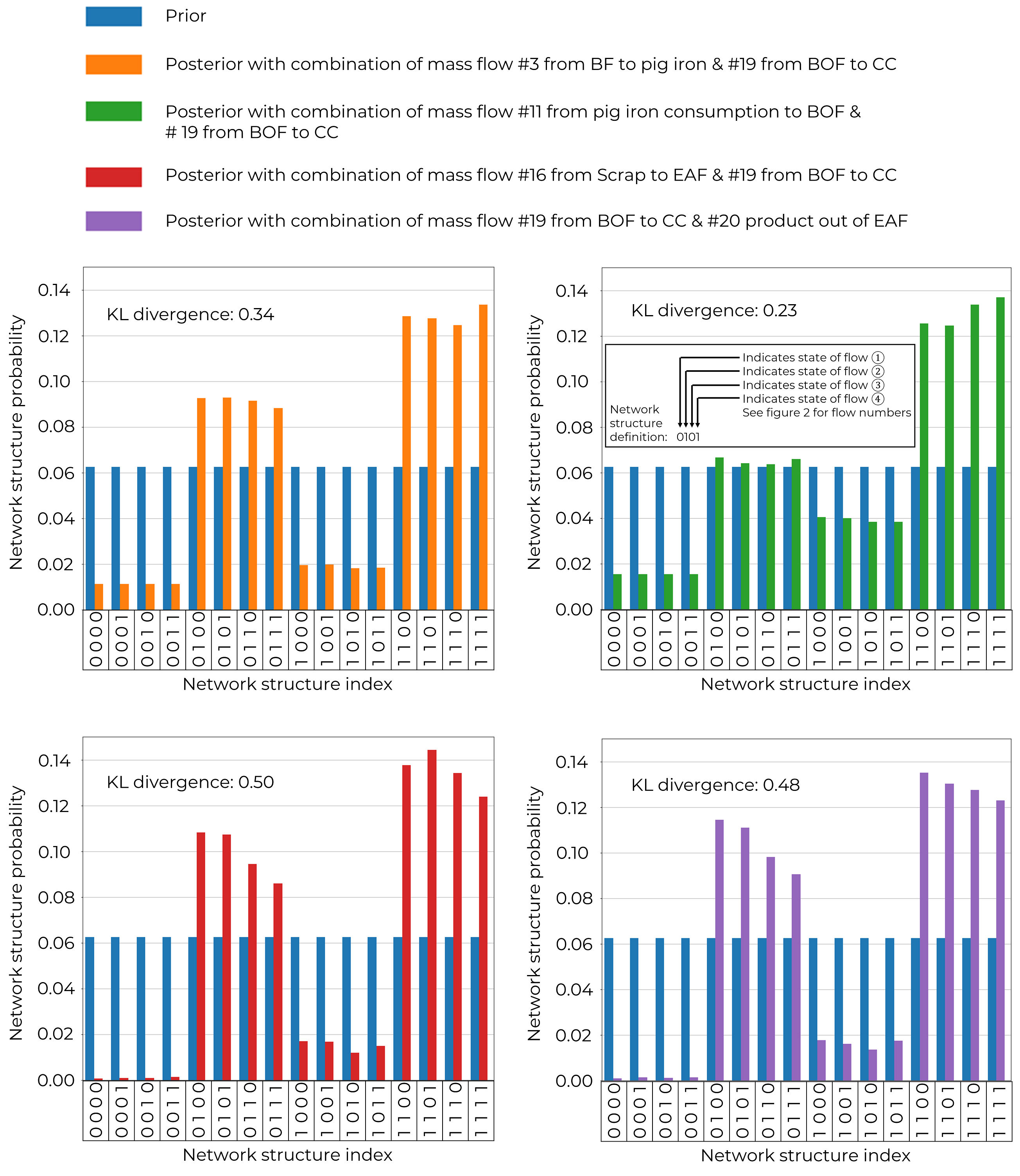}
    \caption{Prior and posterior probability for all 16 candidate network structures with selected data collection on multiple mass flows. Underlying data are available at data repository: \url{https://doi.org/10.7302/k35m-xz34}
    }
    \label{f:batch_validation}
\end{figure}

To assess the alignment between the expected utility of collecting two data points and the actual reduction in network structure uncertainty, we collected the relevant data from the USGS \cite{USGS12a,USGS12b,USGS12c} and WSA \cite{WSA12} (see Table S2). 
\Cref{f:batch_validation} illustrates the prior and posterior distributions of the network structure, along with the KL divergence between them, after incorporating the data on the selected mass flows. This data was used to update the network structure uncertainty through the procedure outlined in \cref{e:bayes_model}.

\subsubsection{Error associated with the numerical estimators}
\label{sss:benchmark}

To identify which numerical estimator of expected utility minimizes error for a given computational cost, we calculate the standard deviation and root mean square error (RMSE) from 100 evaluations of expected utility for separately collecting data on mass flows \#16 and \#19 (single-piece data collection). These evaluations use the same settings as for the case study. The RMSE metric requires knowledge of the exact expected utility value. Since this cannot be obtained analytically, we instead use a reference value that is estimated by a high-quality $\widehat{U}_2$ with 100,000 samples, with sample reuse.

\begin{figure}
    \centering
    \includegraphics[width=\linewidth]{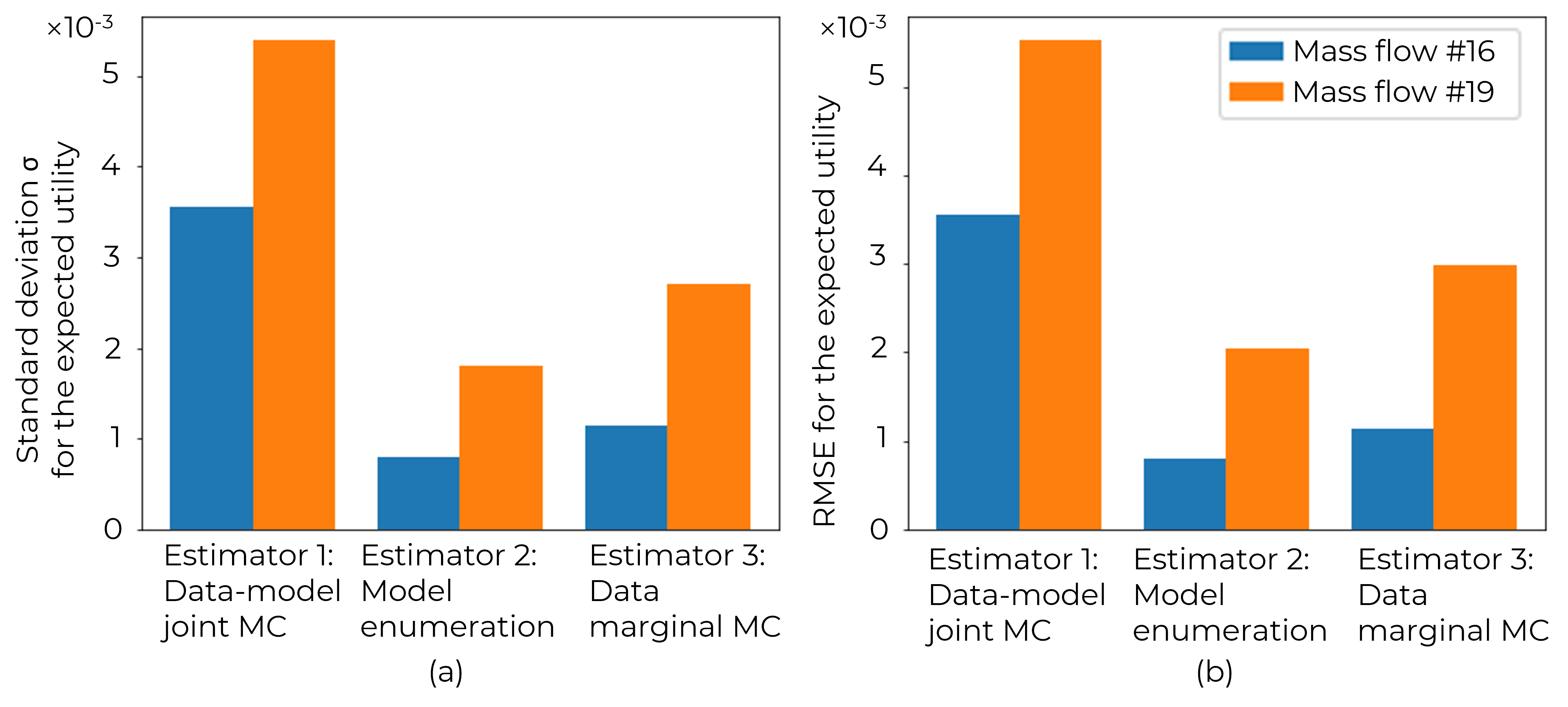}
    \caption{Standard deviation and RMSE from 100 trials of numerical evaluations on expected utility through all 3 estimators with true value generated with Estimator 2): \emph{model enumeration} approach. Underlying data are available at data repository: \url{https://doi.org/10.7302/k35m-xz34}
    }
    \label{f:benchmark}
\end{figure}

\Cref{f:benchmark} presents the error (standard deviation and RMSE) of the numerical estimators, with the ground truth for the RMSE calculation approximated using $\widehat{U}_2$, as it is theoretically the least biased method (see \cref{ss:BOED_framework}). The figure shows that $\widehat{U}_2$ yields the lowest error, exhibiting the smallest standard deviation and RMSE. As demonstrated in S4, 
this result holds true even when a different high-quality numerical estimator with 100,000 samples is used to estimate the true expected utility value.

\section{Discussion}
\label{s:discussion}
In this section, we discuss the implications of shifting from targeting collection of single-pieces of data to batches of data (\cref{ss:discussion_single}), the performance of the numerical estimators (\cref{ss:discussion_benchmark}), and the limitations of the BOED methodology as applied to MFA (\cref{ss:discussion_limitations}).

\subsection{Single-piece versus multi-piece data collection design}
\label{ss:discussion_single}

The case study results show that the hightest expected utility for two-piece data collection (\cref{f:batch_utility}, mass flows \#11 and \#19) is not simply the combination of the highest and second-highest expected utilities for single-piece 
data collection (\cref{f:expected utility}, mass flows \#16 and \#19). This is because the two-piece design problem solved by the BOED framework reveals the synergy of collecting multiple data simultaneously on network structure uncertainty. Intuitively, this can be interpreted as the BOED framework favoring data collection from different subregions of the supply chain. For example, the combination of mass flow \#11 and 19 is more likely to reveal key information about the existence of both connection index 1 (mass flow from scrap to BF) and connection index 2 (mass flow from scrap to BOF) while the combination of mass flow \#16 and 19 is more likely to only provide greater confidence on the existence of targeted connection index 2 as both mass flows \#16 and \#19 share nodes with connection index 2 only.

Comparing the network structure posterior probability results for single-piece (see \cref{f:BOED validation}) and two-piece (see \cref{f:batch_validation}) data collection, it is interesting to note that the actual posterior results for single-piece data collection align well with the expected utility outcomes indicated by the BOED framework. However, the posterior results for two-piece data collection reveal a discrepancy in the order of actual utility achieved, with the highest-to-lowest sequence being \#16 and \#19, \#19 and \#20, \#3 and \#19, and \#11 and \#19, which does not match the expected utility order (highest to lowest: \#11 and \#19, \#3 and \#19, \#16 and \#20, \#19 and \#20). Despite this misalignment, the credibility of the method is not undermined, as expected utility for a given data collection option is inherently an expectation over all possible data values. The credibility of the method stems from its rigorous adherence to mathematical principles (via the laws of probability), rather than any direct comparison to empirical results.

The targeted collection of two pieces of MFA data between rounds of BOED can be easily extended to accommodate the collection of any batch size of MFA data records (see \cref{e:information_gain}). The batch size and the number of BOED rounds are choices for the MFA analyst, as they determine the level of uncertainty in the results that is acceptable, as well as the level of certainty that can be achieved within the constraints of the MFA project's resources.

\subsection{Performance of the numerical estimators}
\label{ss:discussion_benchmark}

\Cref{f:benchmark} shows that, strictly speaking, the second numerical estimator (\emph{model enumeration} estimator) is the most accurate of the three estimators. An MFA analyst would therefore be well justified in using this estimator to evaluate the expected utility of candidates for data collection. However, all three estimators exhibit only small errors, with the standard deviation and RMSE of all estimators being less than 5\% of the nominal expected utility. The absolute value of the bias (i.e., $\sqrt{\text{RMSE}^2-\sigma^2}$) for all three estimators is less than $1\times10^{-3}$.
In the case study, the choice of numerical estimator had minimal impact on the expected utility results, with \cref{f:expected utility} showing that both the absolute expected utility and the order of expected utility from collecting different data records were nearly identical. Therefore, while the $\widehat{U}_2$ is the most accurate numerical estimator, all three estimators are viable options for the BOED framework.

Other considerations, beyond accuracy, when choosing a numerical estimator include computational time. The extended time associated with evaluating $\widehat{U}_2$ in the case study (10 minutes, compared to 30 seconds for evaluating $\widehat{U}_1$ and $\widehat{U}_3$) is due to the necessity to conduct more likelihood evaluations in the $\widehat{U}_2$ method (see \cref{ss:BOED_framework}). In the case study, sample reuse is applied to improve estimator efficiency. However, in cases where sample reuse is not applied, then both $\widehat{U}_2$ \& $\widehat{U}_3$ can be hindered by a smaller sample size, given a fixed computational budget. Therefore, an MFA analyst would also be justified in using Estimator 1 given the improvement in evaluation time for the relatively small loss in accuracy.

\subsection{Limitations}
\label{ss:discussion_limitations}
 
At first glance, the BOED framework introduced in this paper may seem computationally expensive due to the nested Monte Carlo structures in the mutual information estimators. However, it is actually highly efficient, requiring only 30 seconds to 10 minutes on a standard laptop to obtain the results of the case study investigated.

One limitation of the BOED framework in this case study is that it does not account for the varying difficulty and cost associated with collecting data from different sources. A metric reflecting the cost that an MFA analyst is willing to incur per unit of expected information gained (e.g., USD per nat or bit of information) could be incorporated into the optimization problem in \cref{e:optimal_design}.

Another limitation is that the proposed framework is designed to optimally reduce MFA network structure uncertainty but does not explicitly aim to reduce MFA mass flow uncertainty in \cref{e:bayesian_averaged}, which is often the primary objective in MFA. Addressing this would require advancements in \textit{goal-oriented} BOED \cite{Zhong24,Chakraborty24}, capable of targeting uncertainty reduction for more broadly defined QoIs, including specific mass flows in an MFA.

\section{Conclusions and future work}
\label{s:conclusion}

In this study, we have applied a Bayesian optimal experimental design (BOED) framework for targeted collection of MFA data to reduce network structure uncertainty. Since data collection is often a bottleneck in MFA, this work aims to formalize targeted data collection using rigorous probabilistic measures, helping to reduce the time and cost associated with MFA data collection while maximizing the certainty in the final results.

We developed three numerical estimators to solve the data collection optimization problem and demonstrated through a case study that all three methods are viable options for an MFA analyst, with slight trade-offs between speed and accuracy. The case study on the U.S. steel sector highlighted the difference between targeting the collection of single data points versus batches of data, with the optimal choice of data to collect depending on the batch size.

Extensions of this work could involve applying the BOED framework to additional theoretical and real-world MFA case studies to derive next-generation heuristics for targeted data collection. For instance, the case study results suggest the benefits of collecting data on large mass flows that share common nodes with at least one uncertain edge in the network. Another potential extension of this work is the development of goal-oriented BOED methods that target data collection to reduce MFA mass flow uncertainty, accounting for both parametric uncertainty under a fixed network structure and network structure uncertainty.

The BOED framework developed here for reducing MFA network structure uncertainty, and by Liao \textit{et al.} \cite{Liao25b} for reducing MFA parametric uncertainty, could be adapted for use in other areas of industrial ecology. For example, it could inform targeted data collection, measurements, or experiments aimed at reducing parameter uncertainty in gate-to-gate life cycle assessment models of manufacturing processes. Additionally, it could be applied to reduce uncertainty in the model form of environmental process models; e.g., helping discrimination between modeling the per-unit output impacts of an industrial process as having exponential, independent, or logistic dependencies on output volume.


\section{Acknowledgments} 

This material is based upon work supported by the National Science Foundation under Grant No. \#2040013.
The authors declare that they have no conflicts of interest.

\section{Sample Data Availability Statements}

The data that support the findings of this study are openly available in University of Michigan Deep Blue Repositories at \url{https://doi.org/10.7302/k35m-xz34}.

\section{Supporting Information}
This Supporting Information (SI) document includes data, literature reviews and Sankey diagrams for individual candidate network structures helpful to understanding the main article as well as links to Python Scripts and collected MFA data used to conduct the case study.

Please go to this \href{https://deepblue.lib.umich.edu/data/anonymous_link/download/ae4030733ee96335b21f3c25981281b8d740b3d2fccf43041034bb3cf7239aa3?locale=en}{link} (\url{https://doi.org/10.7302/k35m-xz34}):
\begin{itemize}
    \item presentation of the underlying data used to construct the Sankey diagrams in the main paper (Figure 2) in a numerical, tabular format;
    
    \item a Python script for performing BOED to evaluate the expected utility for candidate on mass flow data collection; and

    \item a Python script for performing Bayesian model inference on network structure uncertainty using specified prior PDFs and collected MFA data.
\end{itemize}

\subsection{Parameter priors for the case study}
\label{SI:cs_prior}

In this section, we include the details of the informative priors ($\phi$ and $q$) used for the case study on the 2012 U.S. steel flow. The informative priors for this case study are obtained from expert elicitation through interviews with domain experts. Readers are directed to Dong \textit{et al.} \cite{Dong23} for the details of the methodology to conduct expert elicitation and prior aggregation from multiple experts. The elicitation process conducted is under the assumption of the network structure with all 4 targeted connections existent. As informative priors are applied to both the flows originating from scrap node and continuous cast slab node, adjustment is required to apply these informative priors to candidate network structures where one or more targeted connections are not present in the model. We delete the hyper-parameter(s) of the Dirichlet priors for allocation fractions corresponding to the targeted connection(s) if they do not exist in the candidate network structure, while keeping the rest of the hyper-parameters fixed. For example, the informative prior distribution used for the allocation fractions originating from continuous cast slab to hot strip mill, plate mill, rod and bar mill and section mill when all four connections are present is $\phi\sim Dir(11.46,2.11,2.82,1.81)$. In the case where the connection to the rod and bar mill does not exist, the revised informative prior for this set of allocation fraction will be $\phi\sim Dir(11.46,2.11,1.81)$.

For the exact values of the hyper-parameters used for informative priors, please see the inference code from \href{https://deepblue.lib.umich.edu/data/anonymous_link/download/626ec58fa0d4ce43de584a9f82dca53bb94eaebaadac4eb5168cf1e2fd5c6066?locale=en}{link}.

\subsection{Case study: U.S. steel flow MFA collected data}
\label{SI:MFA_data}

\begin{center}
\begin{longtable}{  m{4em} m{4.5cm} m{4.5cm} m{3cm} m{1.5cm} } 
  \caption{MFA data from 2012.}\\
  \hline
  \textbf{Obs \#} & \textbf{Upstream node} &  \textbf{Downstream node}  & \textbf{Value (Mt)} & \textbf{Source} \\ 
  \hline\hline
  1 & Iron Ore Production &  Export  & 11.2  & 1 \\ 
  \hline
  2 & Iron Ore Consumption &  Blast Furnace  & 46.3  & 1 \\ 
  \hline
  3 & Blast Furnace &  Pig Iron  & 32.1  & 3 \\ 
  \hline
  4 & DRI &  Export  &  0.01 & 2 \\ 
  \hline
  5 & DRI Consumption &  Blast Furnace  & 0.049  & 2 \\ 
  \hline
  6 & DRI Consumption &  Basic Oxygen Furnace  &  1.91 & 2 \\ 
  \hline
  7 & DRI Consumption &  Electric Arc Furnace  &   1.62 &  2\\ 
  \hline
  8 & DRI Consumption &  Cupola Furnace  & 0.01  & 2 \\ 
  \hline
  9 & DRI Consumption &  Other Casting  & 0.01  & 2 \\ 
  \hline
  10 & Pig Iron &  Export  & 0.021  & 2\\ 
  \hline
  11 & Pig Iron Consumption &  Basic Oxygen Furnace  &  31.5  & 2 \\ 
  \hline
  12 & Pig Iron Consumption &  Electric Arc Furnace  &  5.79 & 2 \\ 
  \hline
  13 & Pig Iron Consumption &  Cupola Furnace  &  0.057 & 2 \\ 
  \hline
  14 & Pig Iron Consumption &  Other Casting  & 0.046  &  2\\ 
  \hline
  15 & Scrap Collected &  Export   &  21.4  &  2\\ 
  \hline
  16 & Scrap Consumption &  Electric Arc Furnace  & 50.9  & 2 \\
  \hline
  17 & Scrap Consumption &  Cupola Furnace  &  1.11 & 2 \\ 
  \hline
  18 & Scrap Consumption &  Other Casting  &  0.167 & 2 \\ 
  \hline
  19 & Basic Oxygen Furnace  &  Continuous Casting Slabs  & 36.281  & 4\\
  \hline
  20 & Electric Arc Furnace  &  EAF\_Yield  & 52.414  & 4\\
  \hline
  21 & Pipe Welding Plant &  Pipe and Tubing  & 2.165  & 3\\
  \hline
  22 & Seamless Tube Plant &  Pipe and Tubing  & 2.162 & 3 \\
  \hline
  23 & HSM\_Yield &  Hot Rolled Sheet  & 19.544   & 3 \\ 
  \hline
  24 & CRM\_Yield &  Cold Rolled Sheet  & 11.079  &  3\\ 
  \hline
  25 & Tin Mill &  Tin Mill Products  & 2.009  &  3\\
  \hline
  26 & Galvanized Plant &  Galvanized Sheet  &  16.749 &  3\\
  \hline
  27 & Plate Mill &  Plates  &  9.12 & 3 \\ 
   \hline
  28 & RBM\_Yield &  Reinforcing Bars  & 5.65  & 3 \\ 
  \hline
  29 & RBM\_Yield &  Bars  &  6.7 & 3\\ 
   \hline
  30 & RBM\_Yield &  Wire and Wire Rods  & 2.784  & 3 \\ 
  \hline
  31 & RBM\_Yield &  Light Section  & 2.13  &  3\\ 
  \hline
  32 & SM\_Yield &  Heavy Section  &  5.03 &  3\\ 
  \hline
  33 & SM\_Yield &  Rail and Rail Accessories  & 1.009  & 3 \\
  \hline
  \label{t:SI_2012 data}
 \end{longtable}
\end{center}

\begin{table}[hb]
\begin{tabular}{cc}
 \hline
 \multicolumn{2}{l} {\textbf{Reference in Table 1}} \\
 \hline\hline
 1  & \makecell[l]{USGS. 2012. Iron Ore. Minerals Yearbook.  https://www.usgs.gov/centers/ \\ national-minerals-information-center/iron-ore-statistics-and-information}\\
 \hline
 2 & \makecell[l]{USGS. 2012. Iron and Steel Scrap. Minerals Yearbook. https://www.usgs.gov/centers/\\national-minerals-information-center/iron-and-steel-scrap-statistics-and-information} \\
  \hline
 3 & \makecell[l]{USGS. 2012. Iron and Steel. Minerals Yearbook. https://www.usgs.gov/centers/\\national-minerals-information-center/iron-and-steel-statistics-and-information} \\
  \hline
 4 & \makecell[l]{WorldSteel. 2017. Steel Statistical Yearbook 2017. https://worldsteel.org/steel-by-topic/\\statistics/steel-statistical-yearbook/} \\
  \hline
 \end{tabular}
\end{table}

\subsection{Description on nodes representing the steel flow} 
\label{SI:emission}

\Cref{t:SI_node_emission} presents the description of all nodes featured in the steel flow case study. Some of the nodes in \cref{t:SI_node_emission} are compiler nodes. These nodes are for visualization and calculation purposes. They do not represent actual processes.

\begin{center}
\begin{longtable}{m{17em} m{9cm} }
    \caption{Description of nodes featured in the steel flow case study}\\
     \hline
     Node name  & Note\\
     \hline\hline
     Iron ore production &  Domestic production of iron ore\\
     \hline
     Iron ore consumption  & Compiler node aggregating imported iron ore and domestic iron ore not exported\\
     \hline
     Import iron ore  & \\
     \hline
     DRI production  & \\
     \hline
     DRI & Compiler node describing DRI produced domestically\\
     \hline
     Import DRI & Focus of the analysis is domestic emission: imports assigned 0 emission intensities\\
     \hline
     DRI consumption & Compiler node aggregating imported DRI and domestically produced DRI not exported\\
     \hline
     Blast furnace \\
     \hline
     Import pig iron & \\
     \hline
     Pig iron & Compiler node describing pig iron produced domestically from blast furnace\\
     \hline
     Pig iron consumption & Compiler node aggregating imported pig iron and domestically pig iron not exported\\
     \hline
     Purchased scrap & Post-consumer scrap collected domestically\\
     \hline
     Scrap collected & Compiler node aggregating all post-consumer scrap collected domestically\\
     \hline
     Import scrap & \\
     \hline
     Scrap consumption & Node aggregating post-industrial process scraps and domestically collected post consumer scrap not exported\\
     \hline
     Basic oxygen furnace & \\
     \hline
     Electric Arc furnace & \\
     \hline
     EAF\_yield & Compiler node aggregating all products from electric arc furnace\\
     \hline
     Cupola furnace & \\
     \hline
     Other casting & \\
     \hline
     OC\_yield & Compiler node aggregating all products from other casting process\\
     \hline
     OC\_loss & Compiler node aggregating run-around prep and loss for other casting process\\
     \hline
     Continuous casting - slabs & \\
     \hline
     CC\_yield & Compiler node aggregating all products from continuous cast slabs\\
     \hline
     CC\_loss & Compiler node aggregating run-around prep and loss for continuous cast slab\\
     \hline
     Continuous casting - billets & \\
     \hline
     BT\_yield & Compiler node aggregating all products from continuous cast billets\\
     \hline
     BT\_loss & Compiler node aggregating run-around prep and loss for continuous cast billets\\
     \hline 
     Continuous casting - blooms & \\
     \hline
     BM\_yield & Compiler node aggregating all products from continuous cast blooms\\
     \hline
     BM\_loss & Compiler node aggregating run-around prep and loss for continuous cast blooms\\
     \hline
     Ingot casting & \\
     \hline
     IC\_yield & Compiler node aggregating all products from ingot casting process\\
     \hline
     IC\_loss & Compiler node aggregating run-around prep and loss for ingot casting process\\
     \hline
     Ingot import & \\
     \hline
     Primary mill & \\
     \hline
     PM\_Yield & Compiler node aggregating all products from primary mill\\
     \hline
     Hot strip mill & \\
     \hline
     HSM\_Yield & Compiler node aggregating all products from hot strip mill\\
     \hline
     Plate mill & \\
     \hline
     Rod and bar mill & \\
     \hline
     RBM\_Yield & Compiler node aggregating all products from rod and bar mill\\
     \hline
     Section mill & Profiled rolling process\\
     \hline
     SM\_Yield & Compiler node aggregating all products from section mill\\
     \hline
     Cold rolling mill & \\
     \hline
     CRM\_Yield & Compiler node aggregating all products from cold rolling mill\\
     \hline
     Galvanizing plant & Galvanizing plant taking sheet rolls and coating with zinc\\
     \hline
     Tin mill & Galvanizing plant taking sheet rolls and coating with tin\\
     \hline
     Pipe and tubing & \\
     \hline
     Bars & Cutting process\\
     \hline
     Cold rolled sheet & Manufacturing process of stamping and assembly of steel sheets\\
     \hline
     Galvanized sheet & Manufacturing process of stamping and assembly of steel sheets\\
     \hline
     Hot rolled sheet & Manufacturing process of stamping and assembly of steel sheets\\
     \hline
     Iron product casting & Machining process at critical interfaces of intermediate products\\
     \hline
     Light section & Cutting process\\
     \hline
     Pipe welding plant & \\
     \hline
     Plates & Cutting process\\
     \hline
     Seamless tube plant & \\
     \hline
     Reinforcing bars & Cutting process\\
     \hline
     Rails and rail accessories & Cutting process\\
     \hline
     Heavy section & Cutting process\\
     \hline
     Tin mill products & Manufacturing process of stamping and assembly of steel sheets\\
     \hline
     Wire and wire rods & Forming process to manufacture wire, fasteners and tools\\
     \hline
     Steel product casting & Machining process at critical interfaces of intermediate products\\
     \hline
     Intermediate product import & \\
     \hline
\label{t:SI_node_emission}
\end{longtable}
\end{center}

\subsection{Benchmark with results generated using Estimator 3: \emph{data marginal MC}}

\begin{figure} [h]
    \centering
    \includegraphics[width=\linewidth]{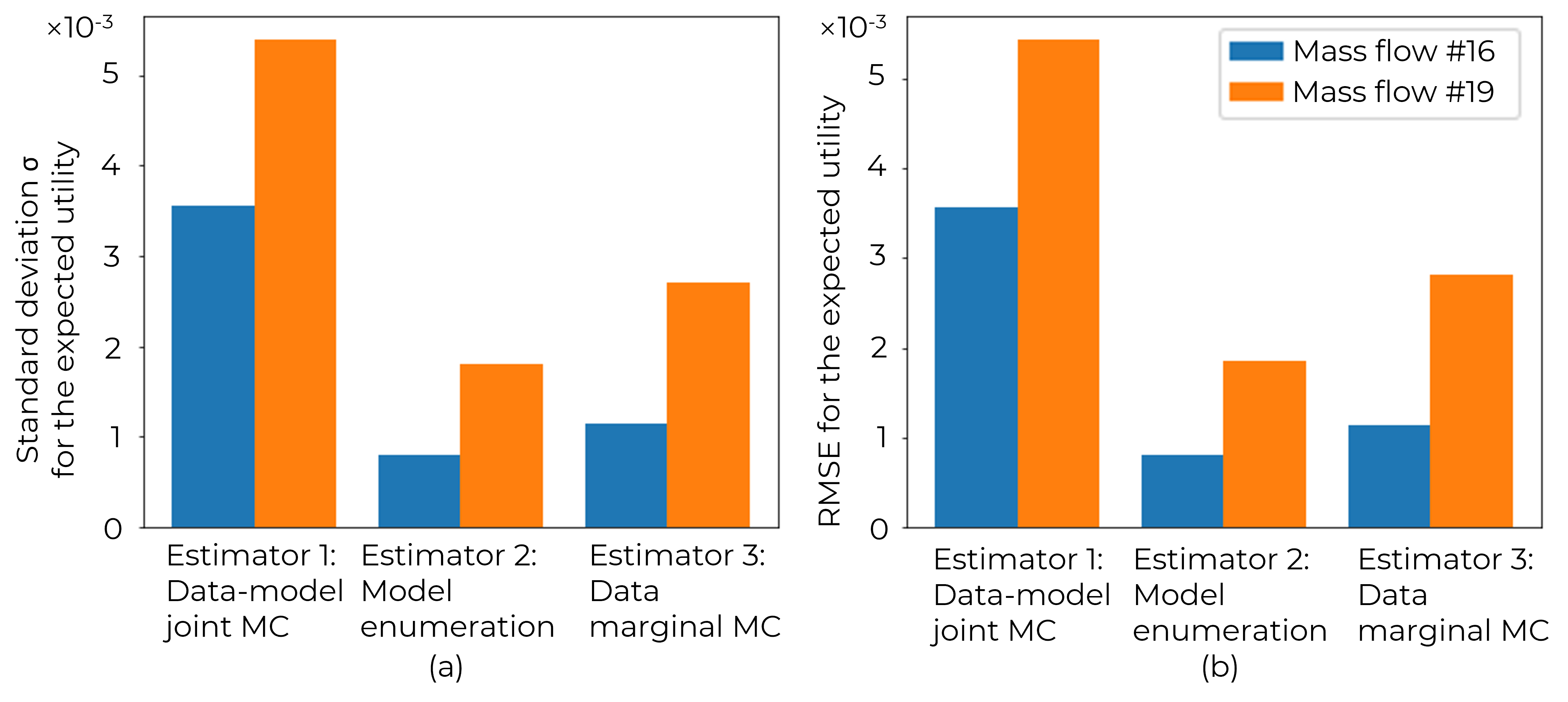}
    \caption{Standard deviation and RMSE from 100 trials of numerical evaluations on expected utility through all 3 estimators with true value generated with Estimator 3: \emph{data marginal MC}}
    \label{f:SI_benchmark}
\end{figure}

\Cref{f:SI_benchmark} shows the benchmark results for all three estimators with the true value generated using Estimator 3: \emph{data marginal MC} with a sample size of 100,000 for both the inner and outer loops.

\bibliography{local}
\bibliographystyle{unsrturl}

\end{document}